\input harvmac
\input graphicx
%
%
\ifx\includegraphics\UnDeFiNeD\message{(NO graphicx.tex, FIGURES WILL BE IGNORED)}
\def\figin#1{\vskip2in}
\else\message{(FIGURES WILL BE INCLUDED)}\def\figin#1{#1}
\fi
\def\Fig#1{Fig.~\the\figno\xdef#1{Fig.~\the\figno}\global\advance\figno
 by1}
%
%
%
%
\def\Ifig#1#2#3#4{
\goodbreak\midinsert
\figin{\centerline{
\includegraphics[width=#4truein]{#3}}}
\narrower\narrower\noindent{\footnotefont
{\bf #1:}  #2\par}
\endinsert
}
%
%
\font\ticp=cmcsc10
\def\undertext#1{$\underline{\smash{\hbox{#1}}}$}

\def\hf{{1\over 2}}
\def\calo{{\cal O}}

\def\cald{{\cal D}}

\def\rtgh{\sqrt\ghat}
\def\ghat{{\hat g}}
\def\Rhat{{\hat R}}

\def\cala{{\cal A}}

\def\thhi{{\hat\theta}_i}
\def\thi{{\hat t}_i}
\def\prodi{\prod_{i=1}^N}
\def\prodij{\prod_{i<j}^N}

\def\matterVev{\left\langle\prod_{i=0}^N\calo_i(x_i)\right\rangle}
\def\subsubsec#1{\noindent{\undertext { #1}}}
\def\mthsu{\mathsurround=0pt  }
\def\leftrightarrowfill{$\mthsu \mathord\leftarrow\mkern-6mu\cleaders
  \hbox{$\mkern-2mu \mathord- \mkern-2mu$}\hfill
  \mkern-6mu\mathord\rightarrow$}
 \def\overleftrightarrow#1{\vbox{\ialign{##\crcr\leftrightarrowfill\crcr\noalign{\kern-1pt\nointerlineskip}$\hfil\displaystyle{#1}\hfil$\crcr}}}
\overfullrule=0pt
%
%

\lref\GMH{
  S.~B.~Giddings, D.~Marolf and J.~B.~Hartle,
  ``Observables in effective gravity,''
  Phys.\ Rev.\ D {\bf 74}, 064018 (2006)
  [arXiv:hep-th/0512200].
}

\lref\dewi{
B. DeWitt, ``The Quantization of Geometry'', in
{\it Gravitation: An Introduction to Current Research},  ed. Witten L (New
York, Wiley, 1962).}

\lref\david{
  F.~David,
  ``Conformal field theories coupled to   2-d gravity in the conformal gauge,''
  Mod.\ Phys.\ Lett.\ A {\bf 3}, 1651 (1988).
}
\lref\DK{
  J.~Distler and H.~Kawai,
  ``Conformal Field Theory And 2-D Quantum Gravity Or Who's Afraid Of Joseph
  Liouville?,''
  Nucl.\ Phys.\ B {\bf 321}, 509 (1989).
}

\lref\RI{C. Rovelli, in Conceptual Problems of Quantum Gravity ed. by
Ashtekar A and Stachel J (Boston: Birkh\"auser, 1991) 141.}

\lref\RII{
  C.~Rovelli,
  ``Quantum Mechanics Without Time: A Model,''
  Phys.\ Rev.\ D {\bf 42}, 2638 (1990).}

\lref\RIII{C. Rovelli, ``Time in quantum gravity: An hypothesis,'' Phys. Rev. D {\bf 43} 442 (1991).}

\lref\maroone{
  D.~Marolf,
  ``Quantum observables and recollapsing dynamics,''
  Class.\ Quant.\ Grav.\  {\bf 12}, 1199 (1995)
  [arXiv:gr-qc/9404053].
}

\lref\Ambj{J.~Ambjorn, K.~N.~Anagnostopoulos, U.~Magnea and G.~Thorleifsson,
  ``Geometrical interpretation of the KPZ exponents,''
  Phys.\ Lett.\  B {\bf 388}, 713 (1996)
  [arXiv:hep-lat/9606012]\semi
J.~Ambjorn and K.~N.~Anagnostopoulos,
  ``Quantum geometry of 2D gravity coupled to unitary matter,''
  Nucl.\ Phys.\  B {\bf 497}, 445 (1997)
  [arXiv:hep-lat/9701006].
}

\lref\marotwo{
  D.~Marolf,
  ``Almost ideal clocks in quantum cosmology: A Brief derivation of time,''
  Class.\ Quant.\ Grav.\  {\bf 12}, 2469 (1995)
  [arXiv:gr-qc/9412016].
}

\lref\GiLione{
  S.~B.~Giddings and M.~Lippert,
  ``Precursors, black holes, and a locality bound,''
  Phys.\ Rev.\ D {\bf 65}, 024006 (2002)
  [arXiv:hep-th/0103231].
}
\lref\GiLitwo{
  S.~B.~Giddings and M.~Lippert,
  ``The information paradox and the locality bound,''
  Phys.\ Rev.\ D {\bf 69}, 124019 (2004)
  [arXiv:hep-th/0402073].
}

\lref\SG{
  S.~B.~Giddings,
  ``(Non)perturbative gravity, nonlocality, and nice slices,''
  Phys.\ Rev.\ D {\bf 74}, 106009 (2006)
  [arXiv:hep-th/0606146].
}

\lref\LQGST{
  S.~B.~Giddings,
  ``Locality in quantum gravity and string theory,''
  Phys.\ Rev.\ D {\bf 74}, 106006 (2006)
  [arXiv:hep-th/0604072].
}

\lref\Seib{
  N.~Seiberg,
  ``Notes on quantum Liouville theory and quantum gravity,''
  Prog.\ Theor.\ Phys.\ Suppl.\  {\bf 102}, 319 (1990).
}

\lref\Polchb{
  J.~Polchinski,
 {\sl String theory. Vol. 1: An introduction to the bosonic string.}
}

\lref\Banks{
  T.~Banks, W.~Fischler and S.~Paban,
 ``Recurrent nightmares?: Measurement theory in de Sitter space,''
  JHEP {\bf 0212}, 062 (2002)
  [arXiv:hep-th/0210160].
}

\lref\Zwie{
  B.~Zwiebach,
  ``Closed string field theory: Quantum action and the B-V master equation,''
  Nucl.\ Phys.\ B {\bf 390}, 33 (1993)
  [arXiv:hep-th/9206084].
}

\lref\Banksds{
T.~Banks,
``Cosmological breaking of supersymmetry or little Lambda goes back to  the future. II,''
arXiv:hep-th/0007146.
}
\lref\Fisch{W. Fischler, unpublished (2000);
W. Fischler, ``Taking de Sitter seriously," Talk given at
Role of Scaling Laws in Physics and Biology (Celebrating
the 60th Birthday of Geoffrey West), Santa Fe, Dec. 2000.}

\Title{\vbox{\baselineskip12pt
\hbox{hep-th/0612191}
\hbox{NSF-KITP-06-128}
}}
{\vbox{\centerline{Relational observables in 2d quantum gravity}
}}
\centerline{{\ticp Michael Gary${}^a$ and}  {\ticp Steven B. Giddings}${}^{a,b}$\footnote{$^\star$}
{Email address:
giddings@physics.ucsb.edu}}
\centerline{${}^a$\sl Department of Physics\footnote{$^\dagger$}{Long-term address.}}
\centerline{and}
\centerline{${}^b$\sl Kavli Institute of Theoretical Physics}
\centerline{\sl University of California}
\centerline{\sl Santa Barbara, CA 93106}
\bigskip
\centerline{\bf Abstract}
Local observation is an important problem both for the foundations of a quantum theory of gravity and for applications to quantum-cosmological problems such as eternal inflation.  While gauge invariant local observables can't be defined, it has been argued that appropriate relational observables approximately reduce to local observables in certain states.  However, quantum mechanics and gravity together imply limitations on the precision of their localization.  Such a relational framework is studied in the context of two-dimensional gravity, where there is a high degree of analytic control.  This example furnishes a concrete example of some of the essential features of relational observables. 

\Date{}

\newsec{Introduction}

An important question for a quantum-mechanical theory of gravity is the nature of its observables.  Specifically, it is of great interest to understand how local observation can be defined, and whether there are any limits to such locality.  There are several reasons for this.  The first is that we are manifestly local observers in a quantum universe.  Indeed, quantum effects become very important in describing many scenarios for the evolution of the early universe.  This is particularly true of eternal inflation scenarios, where we may live in one small bubble among a quantum multitude of different regions.  Attempts to make sense of this picture and our role inevitably rely on statements about local quantities, thus motivating the need for a firmer underpinning to such observation.  Moreover, the issue of locality and local observation may play a central role in formulating a full quantum theory of gravity.  This is in particular suggested by the black hole information paradox, which is a sharp challenge to any attempt to reconcile the principles of quantum mechanics with those of gravity.  But, without some notion of locality,
in a strong sense there {\it is} no black hole information paradox; conversely, it is widely believed that some breakdown of the locality of effective field theory is responsible for the resolution of the paradox.

Thus to begin addressing some important problems in gravity one would like an understanding of local observation.  However, as we will review, diffeomorphism invariance prevents the existence of local observables.  There is a clear tension between this and the fact that all we can truly observe is localized -- we have no access to infinity.  A proposed resolution of this is the notion of a {\it relational observable}.   In essence, localization of an observation must be performed in relation to some background state (us,  our detector, the planet, {\it etc.}).  A very interesting consequence of this is that there are intrinsic limits, from both quantum and gravitational effects, on the precise localization of any such observation.  If, as one might expect, no other notion of local observables exists, this strongly suggests intrinsic limits to the very notion of locality.  

Many of these points were outlined in a recent paper\refs{\GMH}.  However, a precise example requires control over the quantum theory, which goes beyond the limits of present attempts to quantize gravity, such as string theory.  But, as pointed out in \GMH, one example with a number of the essential features is a two-dimensional theory of gravity.  In two-dimensions the dynamics of gravity is greatly simplified; at the classical level it is trivial.  Many aspects of such a diffeomorphism-invariant theory are well understood; in particular, when 2d gravity is coupled to 25 free matter fields, the result is a theory indistinguishable from the world-sheet theory of the critical bosonic string.  Thus one should be able to construct and study relational observables there.

The purpose of this paper is to carefully explain these points, and thus construct a non-trivial example of quantum relational observables and some of their properties.  The next section begins with a basic review of some of the ideas of relational observables.  Section three then describes the basic features of two-dimensional gravity, and explains how to formulate relational observables in this theory.  Section four explicitly computes correlators of such relational observables.\foot{For previous discussion of a different kind of relational observable in 2d dynamical triangulations, see \Ambj.} We find that these correlators reduce, in an approximation, to correlators of local observables in two-dimensional spacetime.  However, the failure of this localization to be precise is an example of the quantum limitations mentioned above.  We conclude with  some discussion.

\newsec{Quantum relational observables}

In quantum field theory, local observables are local gauge-invariant combinations of the basic fields.  The problem in the context of gravity is diffeomorphism invariance.  Given a local scalar operator $\calo(x)$, a diffeomorphism with parameter $\xi^\mu$ acts on it as
\eqn\diffact{\delta_\xi \calo(x) = \xi^\mu \partial_\mu \calo(x)\ .}
Thus no local operator is invariant under the gauge symmetry of gravity.  

This produces an important tension -- the observations we make are manifestly approximately local, and certainly physics should supply a mathematical description of what we observe.

A proposed resolution of this conundrum is the notion of a relational observable.  The idea that observation is relative to the observer certainly traces back to Einstein; classical relational observables were long-ago written down by DeWitt\refs{\dewi}, and have more recently been studied in simple one-dimensional examples in \refs{\RII\RI\RIII\maroone-\marotwo}.  More recently, \GMH\ outlined the construction of quantum relational observables in non-trivial field theories.

Relational observables define localization with respect to features of a background state.  This background state can be taken to be diffeomorphism invariant if in particular it satisfies the Wheeler-deWitt equation.  Moreover, the relational observables also can be defined to be diffeomorphism invariant, for example by integrating scalar operators over spacetime to yield invariants under \diffact.

The key is to recover from such a framework some approximate notion of a local observable.  The work of \GMH\ described some of the features needed to approximately derive local observables, and outlined some examples.  One of these is the Z-model.  In this model one considers some background scalar fields $Z^i$, in a quantum state such that in some region they evolve linearly with some coordinate.  Then, schematically, one defines a relational observable for another scalar field $\phi(x)$ by the expression
\eqn\zobs{\calo_\xi =\int d^4 x \sqrt{-g} \phi(x) \prod_i \delta(Z^i(x)- \xi^i)\, }
for given parameters $\xi^i$.  The delta functions would classically isolate $\phi$ at a definite point.  This expression is classically diffeomorphism invariant, but must be renormalized in the quantum theory.  Some general aspects of the quantum definition of such an operator were described in \GMH, and  it was argued that the fact that $Z^i$, $\phi$, and $g$ are quantum fields prohibits a precise localization at a point, and places other limitations on localization.  Those arising due to gravitational physics are particularly of interest, and suggest a gravitational nonlocality principle; see the locality bound of refs.~\refs{\GiLione,\GiLitwo,\SG} as well as its multiparticle generalization\refs{\LQGST} which is related to instrumentation limits on the degree to which one can localize to a collection of points in a region of spacetime\GMH.

This example of the general framework illustrates various essential features.  In particular the critical ingredients to recovering local operators from such a relational framework are the specification of certain kinds of {\it states} and {\it operators}.  Only in specific cases does one recover local observables from more general diffeomorphism-invariant amplitudes; for example in a more general state the expectation value of $\calo_\xi$ might be defined, but not localize.  Moreover, the recovery of a local observable has inherent limitiations arising from both quantum and gravitational effects, some of which were outlined in \GMH.  In short, in this framework, locality is both {\it relative} and {\it approximate}.

In defining operators such as \zobs\ in four or more dimensions, a difficult point is proper regularization and renormalization.  Without gravity, this could for example be accomplished through a cutoff prescription.  Gravity can to a limited extent be treated this way as well, but that clearly leaves some important questions unanswered and in particular one loses control in the planckian domain.    While a number of the properties of such operators are expected based on very general properties of quantum mechanics and gravity, and thus are expected to be robust, it would clearly be helpful to have a precise formulation of such theories and their observables.  And indeed, one expects that if the limits on locality are intrinsic and unavoidable, these limitations should ultimately be a fundamental aspect of any such formulation.  

One way to gain sharper insight on these questions is via their investigation in contexts where we have greater control over the quantum dynamics.  An example is two-dimensional quantum gravity, which is one of the few non-trivial diffeomorphism-invariant theories where we have analytic control.  Particularly important in this context is the connection to perturbative string theory, which is of course treated as a diffeomorphism-invariant theory on the world sheet, and is very well understood.  We will explore this connection in the subsequent sections, and find a controlled formulation of relational observables in that context.  This will, moreover, illustrate some of the constraints on locality outlined in \GMH.

\newsec{The two-dimensional model}

\subsec{Review of 2d gravity}

Two-dimensional gravity is a simple yet non-trivial 
diffeomorphism-invariant theory\refs{\david\DK-\Seib}.  While we are ultimately most interested in its lorentzian formulation,  it can also be defined via the euclidean functional integral
\eqn\ppi{Z = \int {\cald g \cald m\over {\rm Vol(Diff)}} \, e^{-S[m,g]} \ ,}
for a given matter system, with $m$ the matter degrees of freedom.  Here
 $g$ denotes the 2d metric,  and $S[m,g]$ the matter action.  For the example of a free scalar field $X$, in standard string-theory notation,
\eqn\polyact{S[X,g]= {1\over 4\pi\alpha'} \int d^2 x\sqrt g g^{ab} \partial_a X\partial_b X\ ,}
where the characteristic string length is of order $\sqrt{\alpha'}$.  The Einstein action is trivial and omitted.  One must divide by the volume of the gauge group, here diffeomorphisms, Vol(Diff).  Locally the metric is diffeomorphic to a Weyl rescaling of the trivial metric,
\eqn\confgauge{ds^2 = e^\phi \eta_{ab} dx^a dx^b\ }
(with $a=0,1$);
on a higher genus surface diffeomorphisms and Weyl rescalings reduce the general metric to one of a family of metrics, $\hat g(\tau^i)$, parametrized by a finite number of moduli $\tau^i$.  

In a standard procedure, one thus reduces the integral over metrics to one over Weyl factors and moduli, together with a gauge-fixing determinant.  The measure for the integrals  over $m$, $\phi$ and the determinant have implicit dependence on the Weyl factor, which is described by the Liouville action
\eqn\liouact{S_L[\phi,\ghat] = \int_\Sigma d^2x \rtgh \left(\hf 
\ghat^{ab}\del_a
\phi \del_b \phi + \Rhat \phi\right)  \ .}
For conformal matter $m$ with conformal anomaly $c$, The amplitude \ppi\ then reduces to
\eqn\poly{ Z = \int {d_{\ghat}\tau\, {\rm det}_{\ghat} P} \int \cald_\ghat \phi  \cald_{\hat g}m
e^{[(c-25)/48\pi]S_L[\phi,\ghat]}  e^{-S[m,\ghat]}} 
with contributions $c$  and $+1$ contributed to the coefficient from the matter and $\phi$ measures, and $-26$ from the 
gauge determinant $\det P$; the remaining measures are defined using $\hat g$, as explictly indicated.  This provides a derivation for formulas found in \refs{\david,\DK}.

The scalar field with normalization as in \polyact\ can be defined as 
\eqn\cnorm{{\hat X} = \sqrt{\alpha'{25-c\over 24}} \phi\ .}
Performing this rescaling on \poly, one finds that in the special case $c=25$, the Liouville theory simplifies to a free theory.  For simplicity we will largely focus on this case.

One can generalize \poly\ to calculate diffeomorphism-invariant correlators of operators.  A given matter operator $\Phi_i$ with scaling dimension $\Delta_i$ gets ``dressed" with an anomalous $\phi$ dependence,
\eqn\Dress{ \Psi_i = \Phi_i\, e^{A_i\phi}\ .}
The operator
\eqn\opint{\int d^2 x \rtgh \Psi_i}
will then be diffeomorphism invariant for $A_i$ given by 
\eqn\avalues{ A_i^{\pm} = {25-c\over 12} \left[ 1\pm \sqrt{ 1 + 
{24\over 25-c} (\Delta_i
-1)} \right]\ .}
In the case $c=25$, the gravitationally-dressed operator becomes
\eqn\tfdress{\Psi_i = \Phi_ie^{\pm 2\sqrt{\Delta_i -1} {\hat X}/\sqrt{\alpha'}}\ .}

\subsec{2d relational observables}

 Let us assume that the matter system consists of two free scalar fields $X^0$ and $X^1$, and some other matter with total central charge $c'$.  In the case where this matter is 23 free scalar fields, the world-sheet dynamics is identical to that of the critical bosonic string.  It will prove easiest to continue to lorentzian signature gravity on $S^1$, with world-sheet coordinates $-\infty<t<\infty$ and $0<\theta< 2\pi$.

Following the discussion of section two, local observables should emerge in an approximation from correlators of certain diffeomorphism-invariant relational observables in certain states.  Let us first describe the states.  These should satisfy the Wheeler-DeWitt equation, which in the present two-dimensional framework corresponds to the Virasoro constraints.  

Specifically, in analogy to the Z-model of \refs{\GMH}, we want a background configuration with large field momenta.  Classically, we can take
\eqn\classback{ X^0=p^0 t\ ,\ X^1 = R \theta\ ,}
where the field $X^1$ is periodically identified, $X^1\simeq X^1 + 2\pi R$.  In string language, this is a configuration with target-space energy $p^0$ and winding number $w=1$ on a circle of radius $R$.  Physical states, annihilated by the Virasoro generators, with these momenta are easily constructed; see {\it e.g.} \refs{\Polchb}.  In particular, the Virasoro constraint $L_0=1$ implies that for the ground state of $X$ oscillators and the other matter,
\eqn\backshell{(p^0)^2 = {w^2 R^2\over 4} - 2\ ,}
where in conventional string nomenclature we have chosen $\alpha'=2$.  We label this state $|p^0; w\rangle$.

Relational observables are readily constructed.  Begin with a local operator ${\cal O}_i(x^a)$ of the $c'$ matter system, together with ``momenta" $k_i^a = (k_i^0,k_i^1)$ and ${\hat k}_i$, and define
\eqn\tdrel{\calo_i(k_i^a; x^a ) = \exp\{ik^a_i X_a + i{\hat k_i} {\hat X}\} {\cal O}_i(x^a)\ .}
According to \opint, \tfdress, the integral
\eqn\integop{{\hat \calo}_i(k_i^a) = \int d^2 x \sqrt{\ghat} \calo_i(k^a_i; x^a)}
is a diffeomorphism invariant operator for
\eqn\khatdef{{\hat k}_i =\sqrt{c-25\over6\alpha'}\left[1\pm\sqrt{1+{24\over25-c}\left(\Delta_i+{\alpha'k_{ia}k_i^a\over4}-1\right)}\right]\ ,}
or, with $c'=23$, 
\eqn\khatdefb{{\hat k}_i = \pm2\sqrt{{1-\Delta_i\over \alpha'}- {k_i^2\over 4}}\ .}
The idea, in analogy to the Z-model of \GMH, is to Fourier-transform this with respect to the $k^a_i$, to localize the operator $\calo_i$ with respect to the background provided by the state $|p^0; w\rangle$.  Specifically, one can define\foot{Note that since the direction $X^1$ is compact, more precisely the integral over $k^1$ should be a sum over quantized momenta. For large $R$ this is nearly a continuum integral. }
\eqn\localizop{ {\hat\calo}_i({\hat x}_i^a) = \int {d^2k_i\over (2\pi)^2} e^{-\sigma^2 [(k_i^0)^2+ (k_i^1)^2] } e^{2ik_i^0 p^0 {\hat t}_i - i k_i^1 R {\hat \theta}_i} {\hat \calo}_i(k_i^a)}
to localize the operator $\calo_i$ at $\hat x_i^a=(\hat t_i,\hat\theta_i)$, with a resolution determined by $\sigma$.

The next section will carefully treat correlators of operators such as \localizop.  The special case $c'=23$ is particularly simple, and there we can explicitly show that the relational observables 
indeed approximately produce correlators of the local operators $\calo_i$ of the matter theory.  We will also find quantum limitations on the localization, given in terms of the field momenta $p^0$, $R$.

\newsec{Localization and limitations}

\subsec{Correlators}

This section will show how correlators of the relational observables \integop, \localizop\ in the background state corresponding to \classback\
approximately reduce to correlators of the local operators $\calo_i(x^a)$.  This is technically simplest for $c'=23$, in which the mathematics is identical to that of the critical bosonic string.  While we will focus on this case, the results should generalize.

We first characterize the states and operators.  The one-dimensional universe is taken to be a circle, with initial state $|p^0,w=1\rangle$ described in the previous section.  If the relation \backshell\ holds, this satisfies the 2d version of the Wheeler-deWitt equation.  In string language, this is the ground state of the string with unit winding in the $X^1$ direction, which is compactified on a circle of radius $R$.  (We work in units $\alpha'=2$.)

The relational observables are integrals of the operators  \integop, for some given operators $\calo_i(x^a)$, $i=0,\cdots, N$, of the $c'=23$ matter theory.  In string language the operators \integop\ thus correspond to vertex operators, integrated over the worldsheet.  

\subsubsec{Kinematics}

Invariance under field translations, $X^a\rightarrow X^a + v^a$, $\hat X \rightarrow \hat X + \hat v$, implies that nonvanishing amplitudes must conserve the conjugate {\it target space} momenta.  
The relational observables have momenta $k_i^\mu = (k_i^a,{\hat k}_i)$.
Moreover, the final state of the one-dimensional universe is taken to have momentum  vector $p^{\mu\prime}=(p^{0\prime},0,\hat p')$.  The momenta  $p^{\mu\prime}$ and $k^\mu_0$ can thus be determined in terms of the other momenta.  Specifically, define the total momentum of the initial state and the operators with $i =1,\cdots,N$,
\eqn\Pdef{P^\mu= p^\mu + \sum_{i=1}^N k_i^\mu\ .}
Target space momentum conservation then implies
\eqn\kone{k_0^1 = -P^1\ .}
One then needs to solve the mass-shell conditions \backshell, \khatdefb\ and the remaining two momentum-conservation conditions for $k_0^0$, $\hat k_0$, $p^{0\prime}$, and $\hat p'$.

The solution is easily found, and is given by
\eqn\momsol{ p^{0\prime} = {P^0 S \pm {\hat P} \sqrt{ (p^0)^2\left[{\hat P}^2 - (P^0)^2\right]+ S^2}\over (P^0)^2 - {\hat P}^2}\ ,}
with
\eqn\Sdef{S=1 + { (p^0)^2 - P^2\over 2}\ .}
$k_0^0$ is then determined by  ``energy" conservation,
\eqn\kincons{k_0^0=p^{0\prime}-P^0\ ,}
%
%
and $\hat p'$ and $\hat k_0$ by the mass-shell conditions
\eqn\kinconstwo{
(p^{0\prime})^2 = {R^2\over 4} + ({\hat p}')^2-2}
and \khatdefb.
Expanding at large $R$ and thus $p^0$, one finds the expected result
\eqn\approxkin{ p^{0\prime} = p^0+\calo\left({{1}\over{p^0}}\right)\ .}
%

\subsubsec{Lorentzian correlators}

Thus, the goal is to compute correlators of the form 
\eqn\tcorr{
\langle p';w=1|\prod_i \hat \calo_i(k_i^a)  |p; w=1\rangle\ .}
  Having done so, their Fourier transform with respect to the $k_i^a$, as in \localizop, is expected to give  correlators of operators $\calo_i$ localized in two-dimensional spacetime.

A correlator of the form \tcorr\ can be thought of as arising as a functional integral of the form \ppi\ with the topology of the sphere and $N+3$ operator insertions.  As is familiar from string theory, the conformal symmetry allows us to fix the positions of three of the operators, which we take to correspond to the initial and final states in \tcorr, as well as $\calo_{0}$.  The amplitude can then be straightforwardly gauge fixed and computed in terms of Green functions on the sphere.  However, we most easily obtain a tractable result by working in a {\it lorentzian} two-dimensional spacetime.  This can be obtained by conformally mapping the sphere minus the two external-state points to the cylinder, and then analytically continuing to lorentzian time.  

While the correlators \tcorr\ can  be computed directly in terms of the Green functions on the lorentzian cylinder, to make manifest contact with string theory formalism, we infer them from the well-known form of the Virasoro-Shapiro amplitude for four-tachyon scattering in $c=26$ string theory.  This is the correlator of four operators of the form
\eqn\tachop{V(k) = e^{i k\cdot X}\ }
with $k^2=2$.  (For further details, see {\it e.g.} \Polchb.)
After gauge-fixing and in particular taking three of the operators to lie at the points $z=(0,1,\infty)$, the amplitude becomes
\eqn\vsamp{ {\cal A}_{VS}(k_i)= \int d^2 z |z|^{2k_1\cdot k_2} |1-z|^{2k_2\cdot k_4}\ .}
This expression converges for $k_i\cdot k_j>-1$, but to define it in a region corresponding to physical string scattering,  $k_1\cdot k_2 <-1$, one must analytically continue.  The analog of the continued expression is what we seek for our problem.  This is found by conformally mapping to the cylinder, $z=e^w$, and then performing a Wick rotation:
\eqn\wickr{w = \tau + i \theta \rightarrow i(t +\theta)\ .}
This gives the expression
\eqn\vscont{{\cal A}_{VS}(k_i) = i \int_{-\infty}^\infty dt \int_0^{2\pi} d\theta e^{2it} e^{2it k_1\cdot k_2} \left[\left(1-e^{ix^+}\right)\left(1-e^{ix^-}\right)\right]^{k_2\cdot k_4}\ ,}
where $x^\pm=t\pm\theta.$
This integral is oscillatory as $t\rightarrow\pm\infty$, and thus can be defined (with an $i\epsilon$ convergence factor), and in the physical region, where $k_2\cdot k_4 >-1$, is convergent at $t=\theta=0$.

For the amplitude \tcorr, this can be straightforwardly generalized.  As stated, $x_0$ is fixed to an arbitrary location by world-sheet translation symmetry.  The resulting expression is 
\eqn\amp{\eqalign{
\cala&=\langle p';w=1|\prod_i \hat \calo_i(k_i^a)  |p; w=1\rangle\cr
&=\int\left(\prodi d^2x_ie^{2it_i}\right)\langle p';w=1|\prod_{i=0}^N\calo_i(k_i;x_i^a)|p; w=1\rangle\cr
&=\int\left(\prodi d^2x_ie^{-it_i(2p^0k_i^0-2)+i\theta_iRk_i^1-2i{t_0} p^0k_0^0 + i {\theta_0}Rk_0^1}\right)f(k_i;x_i^a)\matterVev }}
with
\eqn\fequals{
f(k_i;x_i^a)=\prodij\left[(e^{ix_i^+}-e^{ix_j^+})(e^{ix_i^-}-e^{ix_j^-})\right]^{k_i\cdot k_j}\ .}
Moreover, in the small $k$ regime, $|k_i^a|\ll 1$,
\eqn\smallk{
 k_i\cdot k_j=2+\calo\left((k_i^a)^2\right)\ .}
As a result, in analogy to \vscont, the kinematics described above is such that, for  operator 
dimensions $\Delta_i\ll1$, the integral converges at $x_i=x_j$.  The integrals are again oscillatory at $t_i\rightarrow\pm\infty$, and thus well-defined with an $i\epsilon$ prescription.  Furthermore, by eqn.~\smallk\ we can approximate
\eqn\fapprox{f(k_i; x_i)
\approx f(x_i)\ ,}
independent of $k_i^a$, with
\eqn\fdeff{ f(x_i)=\prodij\left[(e^{ix_i^+}-e^{ix_j^+})(e^{ix_i^-}-e^{ix_j^-})\right]^2\ .}

\subsec{Localization on the worldsheet}

Given the lorentzian amplitude \amp, one can now study its localization via the prescription \localizop.  Strictly speaking, the integral over the $k_i^a$ should be restricted to $|k_i^a|\ll1$ to take advantage of the above simplifications.  This is essentially accomplished by taking a sufficiently large gaussian parameter, $\sigma\gg1$.  Moreover, finite $R$ implies that $k_i^1$ are quantized as $n/R$.  However, for large $R$ it is a good approximation to replace $\sum_n \rightarrow R \int dk^1$.  The integral over $k_0^a$ is collapsed by the momentum conservation constraints described above.
For simplicity we use translation invariance to set $x_0^a=0$; otherwise the gaussian phases from \localizop\ should also include $x_0^a$ dependence.
We find 
\eqn\amploc{\eqalign{
\cala_{loc}({\hat t_i},{\hat \theta_i})
&=\int\prodi d^2x_i e^{2it_i}\int_{-\infty}^\infty\left(\prodi{d^2k_i\over(2\pi)^2}e^{-\sigma^2[(k_i^0)^2+(k_i^1)^2]+2ik_i^0p^0\thi-ik_i^1R\thhi}\right)\cr
&\quad\quad\quad\quad\quad\quad\langle p';w=1|\prod_{i=0}^N\calo_i(k_i;x_i) |p; w=1\rangle\cr
&\simeq{1\over (4\pi\sigma^2)^N}\int\left(\prodi d^2x_i e^{2it_i}e^{-{(p^0/\sigma)^2}(t_i-\thi)^2-{(R/2\sigma)^2}(\theta_i-\thhi)^2}\right)f(x_i)\matterVev\ }}
where we drop subleading terms in the $1/R$ and $1/p^0$ expansions.

Thus, up to an overall factor, this expression has reduced to the promised correlator of the operators $\calo_i(x_i)$ of the $c'=23$ matter theory, localized near the points $({\hat t_i},{\hat \theta_i})$ specified in the definition of the relational observables.  The points with $i>0$ are specified relative to an arbitrarily chosen $\hat t_0=\hat \theta_0=0$.  As expected, the localization is not infinitely precise; the resolutions are 
\eqn\resoluts{\Delta t\simeq \sigma/p^0\quad ,\quad \Delta \theta \simeq \sigma/R\ .}

\subsec{Limitations on localization}

As described in our general discussion, relational observables here approximately reduce to local operators, but there are intrinsic limits to the precision to which they are localized.  The present example clearly illustrates this, through equation \resoluts.  Specifically, recall that we only obtain the expression \amploc\ in the limit $\sigma\gg1$; this limits the resolution for fixed background field momenta $p^0, R$.  

Notice that these limitations also make sense in the string-theory interpretation.  The worldsheet statement $\Delta\theta\roughly> 1/R$ translates into the target space statement $\Delta X^1 \roughly> 1$, as illustrated in the figure.  This corresponds to the statement that target-space localization is limited by the string scale.  On the other hand, by taking the circle around which the string wraps to be larger, there is no limit on the worldsheet resolution.

\Ifig{\Fig\figlabel}{As $p^0,R$ are increased with $\ell_s$ held fixed, a string-scale area on the worldsheet becomes smaller relative to the worldsheet area.}{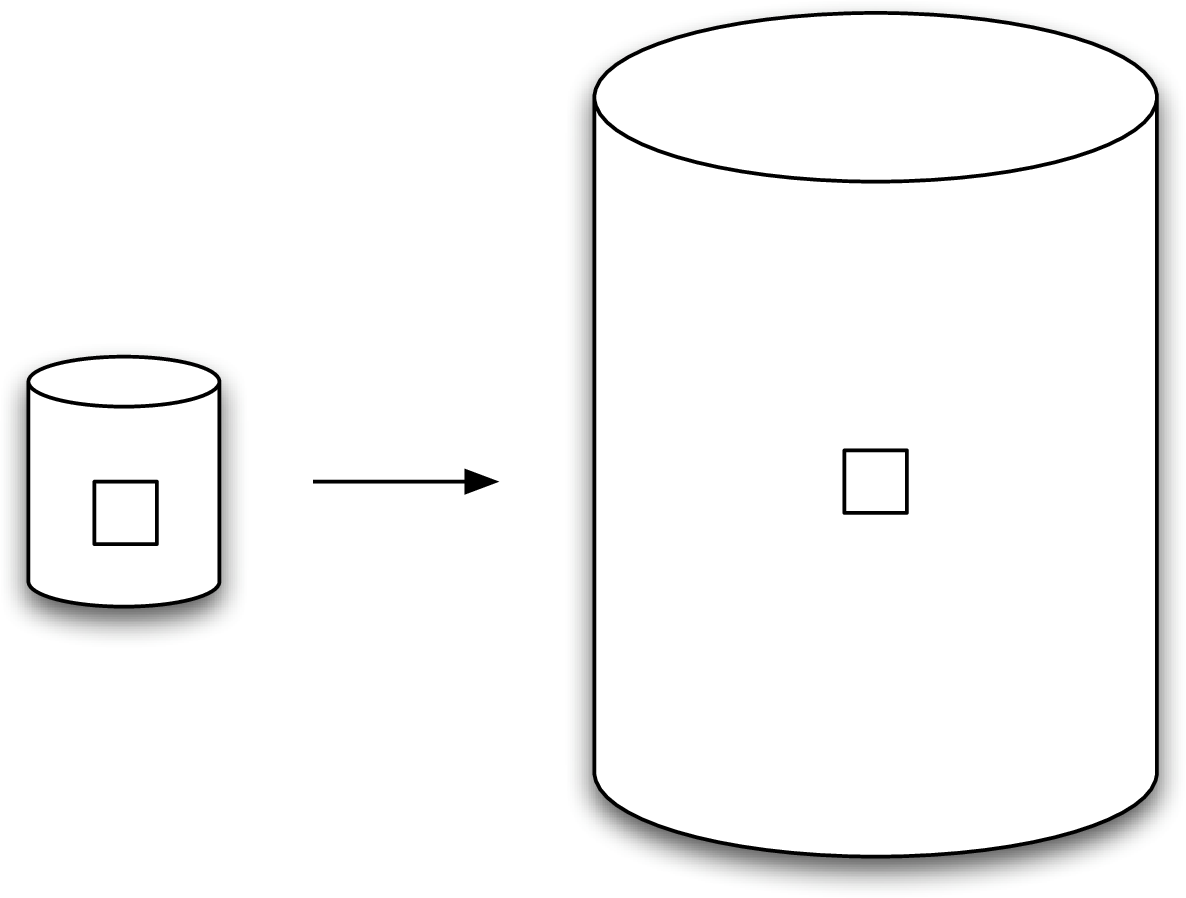}{4}

These do not conform to all of the expectations for higher-dimensional relational observables, but there are reasons for this.  First off, as explained in \GMH, in a theory with a cutoff the resolution to which we can localize is limited by the cutoff.  In the present case there is no such limitation -- the resolution is arbitrarily fine.  This is related to the fact that no cutoff was needed to define the integral \amploc\ for sufficiently small $k_i$ and $\Delta_i$, and appears related to conformal symmetry.  
Secondly, in higher dimensions, the backreaction of the gravitational field limits the size of the field momenta\GMH.  But here, due to the relative triviality of 2d gravity, there is no analogous gravitational limitation on $p^0$ and $R$.

\newsec{Discussion}

The two-dimensional example of this paper furnishes one concrete starting point for thinking about relational observables.  It is interesting to consider both lessons and possible features of the generalization to higher dimensions, amplifying and extending discussion in \GMH.  Ultimately a deeper understanding of these could be important both to shed light on the fundamental quantum dynamics of gravity, and on other important gravitational issues such as the interpretation of eternal inflation.

First, in the present example, amplitudes that correspond to correlators of relational observables can be {\it defined}, whether or not we work in a background such that we approximately recover correlators of local operators.  This is clear upon reconsidering amplitudes of the form \amploc\ where for example the background state doesn't have large momentum/winding.  Thus we have a notion of precise correlators of quantum-mechanical observables that can be {\it computed}, but localization is imprecise and only emerges in an approximation in certain states and for certain operators.  Moreover, a measurement framework for two-dimensional observers would only be expected to arise for certain states, as is further described in \GMH.  These comments for example stand in contrast to discussion in \refs{\Banks}, which suggested inherent imprecision in defining observables.

Of course, as we have noted, the relative triviality of two-dimensional gravity means we don't encounter some of the limits arising in higher dimensional gravity, and it is important to contemplate how the story would differ there.  At long distances the correct theory of gravity is believed to be Einstein's general relativity, but as yet we lack a complete quantum formulation; string theory could be such a formulation, but has not yet answered some fundamental and critical gravitational questions.  Moreover, if the limitations on localization that are apparently emerging are indeed fundamental, as expected, any fundamental formulation should incorporate these limitations.  

Since string theory is a good candidate for a quantum theory of gravity, it is interesting to contemplate how one would formulate relational observables there.  In particular one needs to identify gauge-invariant operators.  The intricate gauge symmetries of string theory are apparently powerful constraints on such operators.  Particularly of interest is the closed string case, which describes gravity.  For example, if one were to start with a closed string field theory formulation, such as in \refs{\Zwie}, one candidate for a relational observable is the derivative of the action with respect to the string coupling, $\partial_g S[\Psi]$.  This gives a gauge-invariant object whose leading term is a cubic expression in string fields.  In certain states, therefore, this should act like a relational observable in analogy to those in the $\psi^2\phi$ model described in \GMH. We hope to explore these points further in the future.

A reasonable viewpoint is that, whatever the fundamental theory is, it is quantum-mechanical, and thus described by a space of states respecting quantum superposition.  Whatever this space is, it is also very reasonable to believe that these states are well described by quantum field theory coupled to general relativity in the low-energy regime, but that it may have no local description on a fundamental level.  In some contexts it may even be finite-dimensional\refs{\Banksds,\Fisch}.  Moreover, it may have no intrinsic definition of time; for example, in the limit where general relativity is valid, we seek solutions of the Wheeler-deWitt equation, ${\cal H} \Psi=0$.
A natural conjecture, extending the discussion of this paper, is that quantum observables can be defined on this space of states, and some of these can be thought of as relational observables.  The ones that are would approximately reduce to local field theory observables in certain states.  Conversely, the limitations on locality that we expect from pushing the limits of our low-energy/long-distance analysis are expected to specifically furnish important constraints on the structure of the space of states.  Some of these restrictions could be summarized by  a gravitational nonlocality principle -- the locality bound\refs{\GiLione,\GiLitwo,\SG}, its multiparticle generalization\LQGST, instrumentation bounds as described in \GMH, and/or other related constraints arising from features of quantum mechanics and gravity.

\bigskip\bigskip\centerline{{\bf Acknowledgments}}\nobreak

We have greatly benefited from conversations with J. Hartle, D. Marolf, and J. Polchinski. This work
was supported in part by the Department of Energy under Contract DE-FG02-91ER40618, and by grant 
RFPI-06-18 from the Foundational Questions Institute (fqxi.org).  SBG also acknowledges  the Kavli Institute for Theoretical Physics and the organizers of the String Phenomenology program both for hospitality and for support under National Science Foundation  Grant No.
PHY99-07949.


\listrefs
\end